\documentclass[nonacm, sigconf]{acmart}

\setcopyright{none}
\renewcommand\footnotetextcopyrightpermission[1]{}

\usepackage{balance}

\begin{document}

\title{Serverless End Game:  Disaggregation enabling Transparency }

\author{Pedro Garcia Lopez}
\affiliation{
	\institution{Universitat Rovira i Virgili/IBM Watson Research, Spain/USA}
}
\email{pedro.garcia@urv.cat}

\author{Aleksander Slominski}
\affiliation{
		\institution{IBM Watson Research, USA}
}
\email{aslom@us.ibm.com }

\author{Simon Shillaker}
\affiliation{
	\institution{Imperial College London, UK}
}
\email{s.shillaker17@imperial.ac.uk}

\author{Michael Behrendt}
\affiliation{
	\institution{IBM Deutschland Research \& Development GmbH, DE}
}
\email{michaelbehrendt@de.ibm.com}

\author{Bernard Metzler}
\affiliation{
	\institution{IBM Zurich, CH}
}
\email{bmt@zurich.ibm.com}

 \date{}

\begin{abstract}
For many years, the distributed systems community has struggled to smooth the transition from local to remote computing. Transparency means concealing the complexities of distributed programming like remote locations, failures or scaling. For us, full transparency implies that we can compile, debug and run unmodified single-machine code over effectively unlimited compute, storage, and memory resources.

We elaborate in this article why resource disaggregation in serverless computing is the definitive catalyst to enable full transparency in the Cloud. We demonstrate with two experiments that we can achieve  transparency today over disaggregated serverless resources and obtain comparable performance to local executions.  We also show that locality cannot be neglected for many problems and we present five open research challenges: granular middleware and locality, memory disaggregation, virtualization, elastic programming models, and optimized deployment.

If full transparency is possible, who needs explicit use of middleware if you can treat remote entities as local ones?  
Can we close the curtains of distributed systems complexity for the majority of users?

\end{abstract}


\maketitle

\section{Introduction}
Transparency is an archetypal challenge in distributed systems that has not yet been adequately solved. Transparency implies the concealment from the user and the application programmer of the complexities of distributed systems. Colouris et al. \cite{colouris} define eight forms of transparency: access, location, concurrency, replication, failure, mobility, performance, and scalability.

But, despite all previous efforts, the problem is still open as seen in recent literature. For example, as stated in \cite{occupy}: ``Our proposal in this paper was motivated by a professor of computer graphics at UC Berkeley asking us: ``Why is there no cloud button?  ``He outlined how his students simply wish they could easily ``push a button`` and have their code (existing, optimized, single-machine code)  running on the cloud``.


Waldo et al. \cite{waldo} explain that the goal  of merging  the programming and computational models of local and remote computing is not new. They state that around every ten years ``a furious bout of language and protocol design takes place and a new distributed computing paradigm is announced``. They mention messages in the 70s, RPCs in the 80s, and objects in the 90s.

In every iteration, a new wave of software modernization is generated, and applications are ported to the  newest and hot paradigm.  Waldo et al. claim that all these iterations may be evolutionary stages to unify  both local and distributed computing. But they are pessimistic, and they believe that this will not be possible because of latency, memory access, concurrency and partial failure. 

This visionary paper even considers that in the future hardware improvements could make the difference in latency irrelevant,  and that differences between local and remote memory could be masked.  But they still claim that concurrency and partial failures preclude the unification of local and remote computing.   Unlike an OS, they are telling us that a distributed system has no single point of resource allocation, synchronization, or failure.

But, what if novel cloud technologies could  make the unification of local and remote paradigms possible?  Are we close to the end of the cycles of software modernization?
Can we just compile to the Serverless SuperComputer \cite{timwagner}?

This paper argues that recent reductions in network latency \cite{latency,attack} are boosting resource disaggregation in the Cloud, which is the definitive catalyst to achieve transparency.  Even if existing Cloud services are still in the millisecond range (100ms Lambda overhead, 10ms in Kafka, 5-20ms in S3), disaggregation has already fueled the creation of serverless computing services like Function as a Service, Cloud Object Storage, and messaging. If we can go down to  {\textmu}s RPCs \cite{rpc1,rpc2}, novel opportunities for transparency will emerge \cite{granular, attack}. 


We present two experiments that analyze the trade-offs between performance and cost when comparing serverless and local computing resources. Even if it seems counter-intuitive,  we will also justify that disaggregation does not make locality irrelevant for many problems and it cannot be ignored.  

The Serverless End Game (enabling transparency)  will arrive when all computing resources (compute, storage, memory) can be offered in a disaggregated way with unlimited flexible scaling.  This will also require a new generation of locality-aware scalable stateful services, smartly combining disaggregation and local resources. We finally study five open research challenges required to achieve full transparency for most applications: (i) granular middleware and locality, (ii) memory disaggregation, (iii) virtualization, (iv) elastic programming models, and (v) optimized deployment.

\section{DDC path to  transparency}

The DDC path is probably the more direct but also the more shocking for the distributed systems community. 
In line with recent industrial trends on Disaggregated Data centers (DDC) \cite{disaggregation}, it implies a distributed
OS transparently leveraging disaggregated hardware resources like processing, memory or storage.

A canonical example is LegoOS: A disseminated, distributed {OS} for hardware resource disaggregation \cite{legoos}. 
LegoOS exposes a distributed set of virtual nodes (vNode) to users. Each vNode is like a  virtual  machine managing its own disaggregated processing, memory and storage resources.   LegoOS achieves transparency and backwards compatibility by  supporting  the  Linux  system  call  interface and  Linux ABIs (Application Binary Interface), so that existing unmodified Linux applications can run on top of it.  Even  distributed  applications that run on Linux can seamlessly run on a LegoOS cluster by running on a set of vNodes. For example, LegoOS shows how two unmodified applications can be run in a distributed way: Phoenix (a single-node multi-threaded implementation of MapReduce)  and TensorFlow.

Another relevant work is Arrakis: The Operating System is the Control Plane \cite{arrakis}. Arrakis comes from previous efforts aimed at optimizing the kernel code paths  to improve data transfer and latency in the OS. In Arrakis, applications have direct access to virtualized I/O devices, which allows most I/O operations to bypass the kernel entirely without compromising process isolation. Arrakis virtualized  control plane approach  allows storage solutions to be integrated with applications, even allowing the development of higher level abstractions like persistent data structures.  Even more, Arrakis control plane is a first step towards integration with a distributed data center network resource allocator.

 If the OS can be extended with unbounded resources in a transparent way, distribution may no longer be needed for many applications -- single-node parallel programming is sufficient.  This is completely in line with the following assessment from the COST paper \cite{COST}: ``You can have a second computer once you’ve shown you know how to use the first one``.  This paper presents a critique of the current research in distributed systems, and even suggests that ``there are numerous examples of scalable algorithms and computational models; one only needs to look back to the parallel computing research of decades past``.

COST stands in that paper for the ``Configuration that Outperforms a Single Thread``. They mainly compare optimized single-threaded versions of graph algorithms, with their equivalents in distributed frameworks like Spark, Naiad, GraphX, Giraph or GraphLab. For example, Naiad  has  a COST of 16 cores for executing PageRank on the twitterrv graph, which means that Naiad needs 16 cores to outperform a single-threaded version of the same algorithm in one machine.

An important reflection from this paper is that the overheads of distributed frameworks (coordination, serialization) can be extremely high just in order to justify scalability. But the COST paper is not proposing a solution to the scalability problem, since it is obvious that a single machine cannot scale enough for many algorithms.

But, what happens if we combine the COST idea with the DDC research? This is precisely what Gao et al.\cite{disaggregation} validated in a simple experiment comparing a COST version with a COST-DDC one that relies on disaggregated memory (Infiniswap \cite{infiniswap}). They demonstrate in this paper that the same code can overcome the memory limits thanks to disaggregation and still obtain good performance results.

DDC is openly challenging the so-called server-centric approach of development for the data center. DDC advocates claim that the monolithic server model where the server is the unit of deployment, operation, and failure is becoming obsolete. They advocate for a paradigm shift where many existing server-centric (cluster computing) approaches must yield to a more efficient way of managing resources in the Data Center. 

The DDC paradigm is presenting  server-centric cluster technologies as obsolete. But current mature Cloud technologies are built on top of  server-centric models with commodity hardware and Ethernet networks. Hardware resource disaggregation is interesting, but it still relies on server-centric clusters for scaling. For example, LegoOS emulates  disaggregated hardware  components  using  a cluster of commodity  servers.   Existing OS approaches like LegoOS have still not dealt with serious distributed systems problems like scalability, consistency and security of the disaggregated resources in a multi-tenant Cloud setting.

\section{Server-centric path to Transparency}

Recent proposals are intercepting language libraries in order to access remote Cloud resources in a transparent way. For example, Crucial \cite{statefulfaas} implements a Serverless Scheduler for the Java Concurrency library. Crucial can run Java threads in Serverless functions transparently, and it also provides synchronization primitives and consistent mutable state data structures over a disaggregated in-memory computing layer.  Crucial does not provide flexible memory scaling or storage transparency, and it is limited to Java applications using that library.

Another example of language level transparency is Fiber \cite{fiber}. Fiber implements an alternative  Python multiprocessing library that works over a scalable Kubernertes cluster. Fiber supports many Python multiprocessing abstractions like Process, Pool, Queue, Pipe and also remote memory in Manager objects. It demonstrates transparency executing unmodified Python applications from the OpenAI Baselines machine learning project. But Fiber does not support transparent disaggregated storage and memory,  and it is limited to Python applications using that library.

The Fabric for Deep Learning (FfDL)  \cite{ffdl} system moves  existing Deep Learning frameworks like PyTorch or TensorFlow to the Cloud on top of cluster technologies like Kubernetes. \cite{ffdl} transparently provides dependability thanks to checkpointing, intercepting storage flows (file system) using optimized storage drivers to cloud object storage,  and supporting locality  with a  gang scheduling algorithm that schedules  all components of a job as a group. 

FfDL gives us two  interesting insights from their authors about the limitations of this system in scalability and transparency. On the one hand,  the scaling was observed to be framework dependent so they could not achieve full scaling transparency. On the other hand, they explain that ``the service was then increasingly used by data scientists who wanted as much control over their FfDL jobs as with their local machine``.  Users wanted to download datasets or Python packages from the public Internet,  interactively debug models, and stream logs to local scripts in order to monitor the progress of jobs. Many of these requests were not possible due to security limitations and the architecture of the system, which frustrated some of their users. 

Another example of transparency in a serverless context is Faasm \cite{faasm}. Faasm exposes a specialised system interface which includes some POSIX syscalls, serverless-specific tasks, and frameworks such as OpenMP and MPI. Faasm transparently intercepts calls to this interface to automatically distribute unmodified applications, and execute existing HPC applications over serverless compute resources. 

Faasm allows colocated functions to share pages of memory and synchronises these pages across hosts to provide distributed state. However, this is done through a custom API where the user must have knowledge of the underlying system, hence breaking full transparency. Furthermore, when functions are widely distributed, this approach exhibits performance similar to traditional distributed shared memory (DSM), which has proven to be poor without hardware support \cite{latencyDSM,munin}.

\section{Limits of disaggregation and transparency}

Current data center networks already enable disk storage disaggregation \cite{locality}, where  reads from local disk are comparable (10ms) to reads from the network.  In contrast, creating a thread in Linux takes about 10 {\textmu}s, still far from the 15ms/100ms (warm/cold) achieved today in FaaS settings. With that, compute disaggregation is already feasible when job time renders these delays negligible.

Advances in datacenter networking and NVMs have reduced access to networked storage to 1 {\textmu}s, however this is still an order of magnitude slower than local memory accesses which are  in the nanosecond range \cite{attack} (100ns), and local cache accesses in the 4ns-30ns range. This means that local memory cannot be neglected, and should be smartly leveraged by memory disaggregation efforts \cite{memory}. Existing efforts in memory disaggregation  \cite{farm, infiniswap, ramcloud, pocket} strive to play in the {\textmu}s range, which can be a limiting factor.

This is directly related to locality and affinity requirements for many stateful applications. The systems community is starting to acknowledge that stateful services need a different programming model and resource management than the stateless ones \cite{granular, onestep}.  Stateful services have very different requirements of coordination, consistency, scalability and fault tolerance, and they need to be addressed differently. Stateful services show the limits of disaggregation versus locality, since in some scenarios locality still matters.

For now, locality still plays a key role in stateful distributed applications.  For example:  (i)  where huge data movements still are a penalty and memory-locality can be still useful to avoid data serialization costs; (ii) where specialized hardware like GPUs must be used \cite{ffdl}; in (iii) some iterative machine-learning algorithms \cite{tangram}; in (iv) simulators, interactive agents or actors\cite{ray}.

Finally, another important limitation is scaling transparency, which means that applications can expand in scale without changes to the system structure or the application algorithms. If the local programming model was designed to use a fixed amount of resources, there is no magic way of transparently achieving scalability, not to mention elasticity. 
Workloads that do not need elasticity, such as enterprise batch jobs or scientific simulations, can use disaggregated resources the same way as local as they do not need scalability. However, for more user driven and interactive services, such as internal enterprise web applications, simple porting of the executables (sometimes referred as ``lift-and-shift``) is rarely enough. The unchanged code is not able to take advantage of the elasticity of disaggregated resources and it is expensive to run code that is not used.

\section{Experiments}

\subsection{Compute disaggregation}

To evaluate the feasibility of compute disaggregation with state of the art cloud technologies, we will compare a compute-intensive algorithm running in local threads in a VM compared to the same algorithm running over serverless functions. We also provide code transparency, since we execute the same code in both cases. To achieve this transparency, we rely on a Java Serverless Executor \cite{statefulfaas} that can execute Java threads over remote Lambda functions. In this case, all state is passed as parameters to the functions/threads, and functions are in warm state, like VMs which are already provisioned.

This experiment runs a Monte Carlo simulation to estimate the value of $\pi$. At each iteration, the algorithm checks if a random point in a 2D square space lies inside the inscribed quadrant. We run 48 billion iterations of the algorithm. For AWS Lambda, the iterations are evenly distributed to 16, 36, 48 or 96 functions with 1792 MB of memory.\footnote{According to AWS documentation, at 1,792MB a function has the equivalent of one full vCPU} For virtual machines, we run a parallel version of the simulation in different instance sizes: c5.4xlarge (16 vCPUs), c5.9xlarge (36 vCPUs), c5.12xlarge (48 vCPUs), c5.24xlarge (96 vCPUs). The algorithm is implemented in Java.

\begin{figure}
\includegraphics[width=\columnwidth]{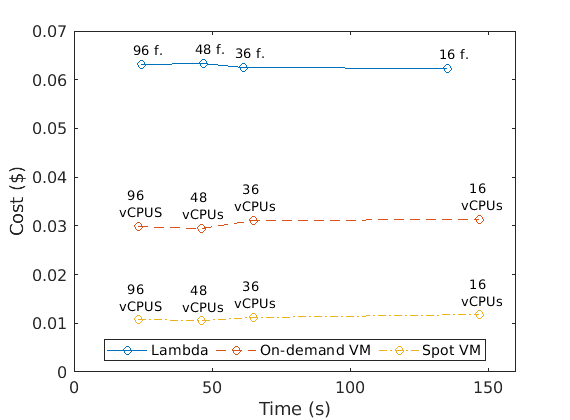}
\caption{Monte Carlo simulation in VMs versus Amazon Lambda Functions}
\label{fig:gups}
\end{figure}

As we can see in Figure~\ref{fig:gups}, the major difference now is cost: for an equivalent execution, disaggregated functions cost 2x more compared to on-demand VMs, and 6x more compared to Spot instances. Surprisingly, computation time is equivalent in the local and remote version using Lambdas. Even considering all the network communication overheads, container management and remote execution, the results for disaggregated computations are already competitive in performance in existing clouds. This is of course happening because this experiment is embarrassingly parallel, and the duration of compute tasks is long enough to make milliseconds (15/100ms) overheads negligible.

\subsection{Memory disaggregation}

The second experiment evaluates the feasibility and costs of both memory and compute disaggregation with existing cloud technologies. In this case, we evaluate a linear algebra algorithm,  Matrix Multiplication (GEMM) which is a good use-case for testing parallel processing on large in-memory data structures. 

We rely on Python frameworks used by data scientists like NumPy and Dask. Dask transparently enables to run the same code  in a single multi-core machine or VM, and in a distributed cluster of nodes. We also compare Dask to a serverless implementation of NumPy called numpywren \cite{numpywren}  using  serverless functions that access data in disaggregated Cloud Object Storage (Amazon S3).

\begin{figure}
\centering
\includegraphics[width=\columnwidth]{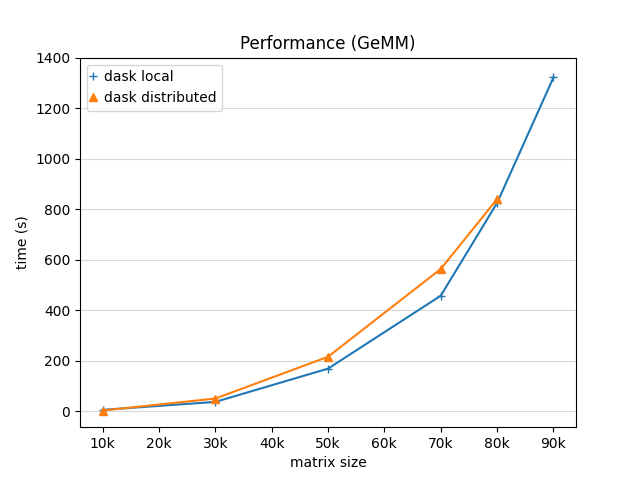}
\caption{Comparing Vertical vs. Horizontal Scaling: GEMM Matrix
 Multiplication in Dask Local vs. Distributed}
\label{fig:gemm2}
\end{figure}

\begin{figure}
\includegraphics[width=\columnwidth]{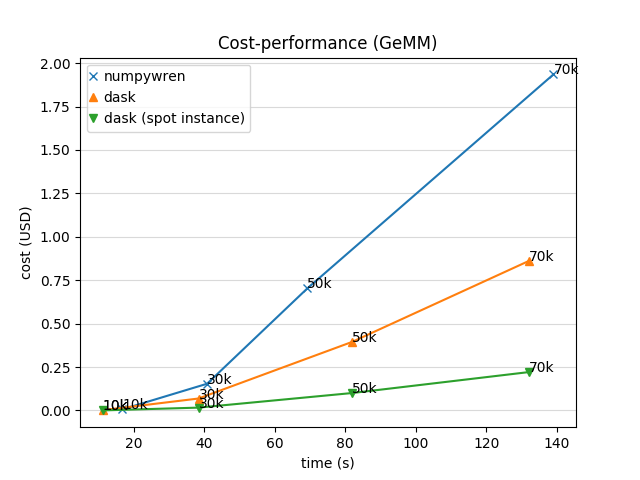}
\caption{Comparing Horizontal Scaling Options: GEMM Matrix Multiplication in Dask Distributed (Spot Instances and on demand VMs) and numpywren (Lambda) for different matrix sizes}
\label{fig:gemm3}
\end{figure}



Our first experiment compares the performance of Matrix Multiplication (GEMM) using Dask in a local VM (1x r5.24xlarge) and in a distributed cluster (6x r5.4xlarge)  using the same resources (96 vCPUs, 768 GiB memory, 10Gb network). Figure~\ref{fig:gemm2} shows that the local version perform slightly better than the distributed one while costing the same.  In this case, locality is avoiding unnecessary data movements and serialization costs, and cluster provisioning. Experiments with 90Kx90K matrices can be executed in the local VM, but not in the equivalent distributed cluster due to resource exhaustion. 

Our second experiment compares the cost and performance of Matrix Multiplication (GEMM) using Dask in a distributed cluster (on demand VMs or Spot instances) and using numpywren over Amazon Lambda and Amazon S3. We calculate compute resources in numpywren (vCPUs) as the ratio between the sum of the duration of every Lambda and the wall-clock time of the experiment. In GeMM (70Kx70K) numpywren uses 553.8 vCPUs and in Dask we use equivalent resources: 552 vCPUs (5x c5.24xlarge, 1x c5.18xlarge).
		
Figure~\ref{fig:gemm3}  shows that Dask obtains the same performance in VMs and Spot instances, but  Spot instances are 4x cheaper than on demand  VMs. numpywren obtains good performance numbers for large matrices, obtaining equivalent performance results for an equivalent Dask cluster in running time. numpywren also shows automatic scaling for any size, whereas the Dask cluster must always be provisioned in advance with the desired amount of resources. Finally, numpywren is much more expensive than the Dask cluster using Spot instances (14x for 10K, 9x for 30K, 6.9x for 50K, 8.7x for 70K).

We see in these experiments what can be achieved today with existing state-of-the-art Cloud infrastructure. Monetary cost is now the strongest reason  for locality in Cloud providers as we see in the pricing models for Lambda, on demand VMs and Spot instances. But even if elastic disaggregated resources are now more expensive, some large scale compute intensive problems like linear algebra are now already  competitive in compute time and scalability.  Further improvements in cloud management control planes and locality-aware placement could reduce costs for elastic resources.






\section{Challenges ahead}



Let us review the major challenges to enable transparency for many applications:

\newpage
\begin{itemize}

\item \textbf{Granular middleware and locality}: In line with granular computing 
\cite{granular,attack}, we require microsecond latencies in existing middleware (compute, storage, memory, communication). In particular, there is a need to handle extremely short instantiation and execution times and more lightweight container technologies. We also require microsecond latencies in disaggregated storage and memory, messaging and collective communication.

Granular applications are amenable to fine-grained elastic scaling, but this will not provide adequate performance without data locality.  Locality and fine-grained resource management may also reduce the current cost of disaggregated resources. Locality is also needed to scale stateful services with different requirements of coordination, concurrency, consistency, distribution, scalability and fault tolerance.  But existing FaaS services provide very limited locality/ affinity mechanisms and limited
networking, precluding inter-function communications. We foresee that next-generation container technologies may enable inter-container communication and provide affinity services for grouping related entities (e.g. gang scheduling \cite{ffdl}).

\item \textbf{Memory disaggregation and Computational memory}:  Disaggregated memory is still an open challenge and there is no available Cloud offering in this line. Many cluster technologies like Apache Spark, Dask, or Apache Ray rely on coupled and difficult-to-scale in-memory storage. Fast disaggregated memory and storage services \cite{farm,pocket} can facilitate the elasticity of many cluster technologies \cite{stuedi2019unification}. 

An important problem here is that disaggregated memory services cannot ignore the memory available in  existing server-centric nodes in most Cloud providers.  One option is to combine both local and remote memory resources efficiently \cite{memory}. Another potential solution here is the recent line on computational memory \cite{nanomemory}  and in-memory computing devices. Compute and memory locality (similar to mammalian brain where memory and processing are deeply interconnected) may considerably enhance computational efficiency.

\item \textbf{Virtualization} Accessing disaggregated resources in a transparent manner requires a form of lightweight, flexible virtualization that does not currently exist. This virtualization must intercept computation and memory management to provide access to disaggregated resources, and must do so with native-like performance and no input from the programmer. Current serverless platforms use Linux containers and VMs for virtualization~\cite{peeking,firecracker}, which have proven to be too heavyweight for fine-grained scaling, and inappropriate for stateful applications~\cite{faasm,granular,onestep,cloudburst}. Software-based virtualization is a more lightweight alternative that is seeing adoption in the serverless context~\cite{microRust,faasm}, and as a replacement for Docker~\cite{krustlet}, but is not yet mature enough to transparently support non-trivial existing applications.

Virtualization necessarily defines an interface between users' code and the underlying system, but the nature of this interface in a transparent disaggregated context is unclear. Exposing a full Linux API makes porting legacy applications easy as shown in LegoOS~\cite{legoos}, but requires heavy engineering and introduces historical idiosyncrasies such as \verb|fork|~\cite{forkInTheRoad}. WASI~\cite{wasi} aims to provide lightweight virtualization with a subset of POSIX-like calls and a custom libc, but can only support a small subset of existing C/C++ applications.  Platform-independent runtimes such as GraalVM~\cite{truffle} raise the virtualization layer into the language runtime itself. This affords flexibility in supporting a range of underlying hardware, but is restricted to a a subset of programming languages and applications.

\item \textbf{Elastic programming models and developer experience:} In some cases, virtualization technologies cannot solve problems like scaling transparency if the  code is programmed to use a fixed amount of resources.  We then need elastic programming models for local machines that can be used without change when running over Cloud resources. Such elastic models should take care of providing the different  transparency types (scaling, failure, replication, location, access) and other aspects of application behavior when it is moved between local and distributed environments.  The local executable APIs may need to be expanded to include elastic programming abstractions for processes, memory, and storage.

To fulfill the vision of disaggregation and transparency it will also be critical to provide tools for developers, enabling them to code both locally and remotely in the same manner with full transparency. Developers will need to be able to use tools to debug, monitor, profile, and if necessary access control planes to optimize their applications for cost and performance.



\item \textbf{Optimized deployment}:  Existing applications are a  blackbox for the cloud, but the transition will imply a ``compile to the Cloud`` process. In this case, the Cloud will have access  over applications' life cycle and it will be able to optimize their execution performance and cost.  This means that they can  perform static analysis to predict resource requirements, dependencies and potential for hardware acceleration. Future Cloud orchestration services will  explicitly leverage  data dependencies and execution requirements for improving workloads and resource management thanks to machine learning techniques \cite{learning, retro}. This compile process will also allow advanced debugging mechanisms for  Cloud applications. 

Transparency efforts for different types of applications will require customizable control planes for applications. Such customization will be based on  advanced observability and fast orchestration mechanisms relying on standard services and protocols.  Monitoring and interception of the different resources (compute, storage, memory, network) should be available and even integrated into the data center, enabling coordinated actuators at different levels. This can enable the creation of millions of tiny control planes \cite{tiny}  adapted to the different applications and programming models.

\end{itemize}

\section{Conclusions}

We argue that full transparency will be possible soon in the Cloud thanks to low latency resource disaggregation. We foresee that next generation serverless technologies may overcome the limitations exposed by Waldo et al. \cite {waldo} more than twenty five years ago. In the next years, we will be able to develop programs without taking care of address spaces, while a modern cloud environment will transparently and efficiently execute those on disaggregated resources.


We demonstrated that it is possible today to transparently leverage remote resources and obtain comparable performance to local deployments. But we also showed that there are still open research challenges to solve in the next years related to  (i) granular middleware and locality, (ii) memory disaggregation, (iii) virtualization, (iv) elastic programming models, and (v) optimized deployment.  

The next frontier for transparency is to go beyond the boundaries of the data center, and  seamlessly support heterogeneous resources in the Cloud Continuum (Hybrid/Edge/Federated/Cloud). Another important challenge is to devise elastic parallel programming models for a single machine that can transparently leverage heterogeneous resources in the Cloud Continuum. 

If transparency is possible soon:  Is this the end of distributed programming for the majority of developers? 
Can we just rely on parallel programming  techniques and be completely oblivious to the underlying distributed infrastructure even for large scale problems?  Who needs explicit use of middleware if you can treat remote entities as local ones? And finally, 
can we close the curtains of distributed systems complexity for the majority of users ?

\begin{acks}
	This work has been partially supported by the EU Horizon2020 programme under grant agreement No 825184. Special thanks to Gerard Paris (Monte Carlo experiment), Pol Roca (GEMM experiment), and Daniel Barcelona for his insights and comments.
\end{acks}

{ \balance
{

\bibliographystyle{ACM-Reference-Format}}
\bibliography{ccr}

\end{document}